# AbGradCon 2021: Lessons in Digital Meetings, International Collaboration, and Interdisciplinarity in Astrobiology

**Tony Z. Jia[1,2,*], Kristin N. Johnson-Finn[1], Osama M. Alian[3], Irene Bonati[1,4], Kosuke Fujishima[1,5], Natalie Grefenstette[2,6], Thilina Heenatigala[1,7], Yamei Li[1], Natsumi Noda[1,8], Petar I. Penev[9], Paula Prondzinsky[1], Harrison B. Smith[1]**

**[1]Earth-Life Science Institute, Tokyo Institute of Technology, 2-12-1-IE-1 Ookayama, Meguro-ku, Tokyo 152-8550, Japan.**

**[2]Blue Marble Space Institute of Science, 1001 4th Ave., Suite 3201, Seattle, Washington, 98154, USA.**

**[3]Michigan State University, 288 Farm Lane, East Lansing, Michigan, 48824, USA**

**[4]Netherlands Institute for Radio Astronomy (ASTRON), Oude Hoogeveensedijk 4 7991 PD Dwingeloo, The Netherlands.**

**[5]Graduate School of Media and Governance, Keio University, 5322 Endo, Fujisawa-shi, Kanagawa 252-0882, Japan**

**[6]Santa Fe Institute, 1399 Hyde Park Road, Santa Fe, New Mexico, 87501, USA**

**[7]National Astronomical Observatory of Japan, 2 Chome-21-1 Osawa, Mitaka, Tokyo 181-8588, Japan.**

**[8]Department of Earth and Planetary Science, The University of Tokyo, 7-3-1 Hongo, Bunkyo-ku, Tokyo 113-0033, Japan.**

**[9]Innovative Genomics Institute, University of California, Berkeley, 2151 Berkeley Way, Berkeley, California, 94720, USA**

***Corresponding Author:**
**Tony Z. Jia**
**Earth-Life Science Institute**
**Tokyo Institute of Technology**
**2-12-1-IE-1 Ookayama, Meguro-ku, Tokyo 152-8550, Japan**
**Tel: +81-03-5734-2708**
**Email: tzjia@elsi.jp**

**Abstract**

The Astrobiology Graduate Conference (AbGradCon) is an annual conference both organized for and by early career researchers, postdoctoral fellows, and students as a way to train the next generation of astrobiologists and develop a robust network of cohorts moving forward. AbGradCon 2021 was held virtually on September 14–17, 2021, hosted by the Earth-Life Science Institute (ELSI) of Tokyo Institute of Technology after postponement of the in-person event in 2020 due to the COVID-19 pandemic. The meeting consisted of presentations by 120 participants from a variety of fields, two keynote speakers, and other career building events and workshops. Here, we report on the organizational and executional aspects of AbGradCon 2021, including the meeting participant demographics, various digital aspects introduced specifically for a virtual edition of the meeting, and the abstract submission and evaluation process. The abstract evaluation process of AbGradCon 2021 is unique in that all evaluations are done by the peers of the applicants, and as astrobiology is inherently a broad discipline, the abstract evaluation process revealed a number of trends related to multidisciplinarity of the astrobiology field. We believe that meetings like AbGradCon can provide a unique opportunity for students and early career researchers in astrobiology to experience community building, inter- and multidisciplinary collaboration, and career training and would be a welcome sight in other fields as well. We hope that this report provides inspiration and a basic roadmap for organizing future conferences in any field with similar goals.

## 1. Introduction

The Astrobiology Graduate Conference (AbGradCon) began in the United States of America in 2004 with the support of the (then named) NASA Astrobiology Institute (National Academies of Sciences, Engineering, and Medicine *et al.*, 2019), and since its inception has been an interdisciplinary meeting for and organized by students, postdoctoral fellows, and other early career astrobiology researchers (McGonigle, Motamedi and Rapf, 2019). Some goals of AbGradCon include: introduction of astrobiology to participants without much astrobiology background, exposure to different astrobiology-related fields through one track of presentations, creation of research networks with other early career astrobiology researchers, and practice giving presentations in a formal conference setting. AbGradCon is often the first official conference presentation experience for many participants. Throughout its history, the meeting has been held annually at rotating host sites (Som *et al.*, 2009) (with one funding-related cancellation in 2006 (Riccardi *et al.*, 2006)), mostly in the continental USA, but non-continental-USA-hosted editions of AbGradCon have been hosted in San Juan, Puerto Rico (2007), Taellberg, Sweden (2010), and Montreal, Canada (2012). In order to increase international participation in AbGradCon, the Earth-Life Science Institute (ELSI) at Tokyo Institute of Technology in Meguro-ku, Tokyo, Japan was chosen as the site of the 2020 AbGradCon meeting (henceforth known as AbGradCon 2020), to have been held on September 14-18, 2020. As such, the major theme of AbGradCon 2020 was to be international collaboration; the conference location in East Asia (along with Japan's various growing research communities in astrobiology and related fields, partially catalyzed by ELSI and various domestic research societies (Scharf *et al.*, 2015; Chan *et al.*, 2019; Jia and Kuruma, 2019; Mariscal *et al.*, 2019)) would allow more ease of travel access for participants from Asia and Oceania, while still allowing connections from various points in the Americas and Europe.

However, during the planning of AbGradCon 2020, the COVID-19 (coronavirus disease 2019) pandemic began (Platto *et al.*, 2021), and AbGradCon 2020 was not held. Instead the focus would be on organizing an edition of the meeting in 2021 (AbGradCon 2021) hosted at ELSI with the same goals, structure, and theme as AbGradCon 2020, to be held on September 14–17, 2021 (**Fig. 1**). Nevertheless, despite some progress in understanding the mechanism of COVID-19 detection (Ji *et al.*, 2020; Kevadiya *et al.*, 2021), spread (Tang *et al.*, 2020), mitigation (Howard *et al.*, 2021; Nande *et al.*, 2021), and vaccine development (Tregoning *et al.*, 2021), the global pandemic situation did not subside by September 2021, necessitating a switch to a virtual format for AbGradCon 2021. The new version of the meeting came with its own perks and challenges. For example, while it would be more accessible to attendees from all around the world without travel, balancing time zones for a wide group of worldwide participants became the new challenge. Since the beginning of the pandemic in early 2020 many conferences and meetings have been held virtually resulting in a phenomenon termed “Zoom fatigue” (Wiederhold, 2020; Bailenson, 2021; Williams, 2021). “Zoom fatigue” can result in less participation and excitement regarding virtual conferences, leading to minimized network building between conference participants; as such, the organizers designed AbGradCon 2021 to mitigate “Zoom fatigue” and encourage participants to develop relationships with other early career participants of the meeting.

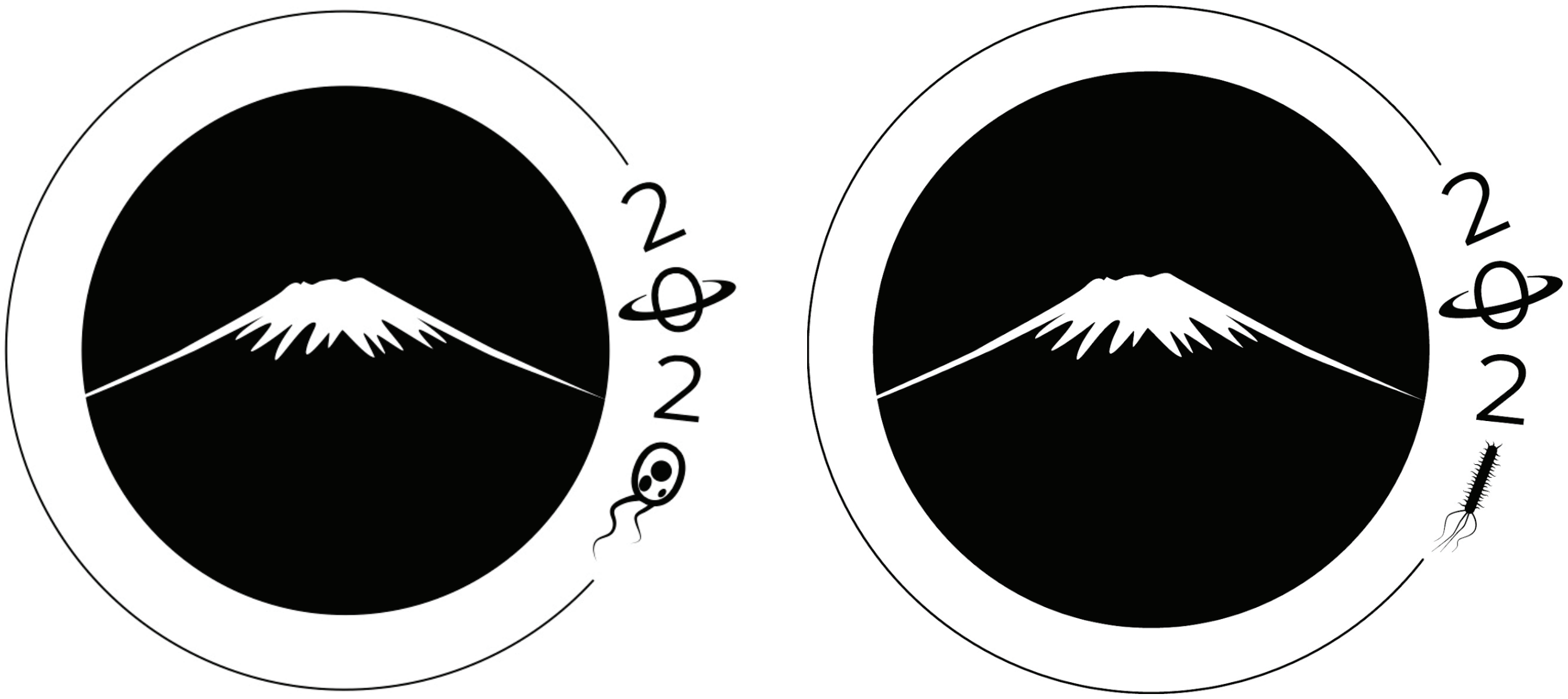

**Figure 1. The AbGradCon 2020 (left) and AbGradCon 2021 (right) conference logos. Designed by Lucy Kwok (http://be.net/thislucykwok).**

Despite the challenges of organizing AbGradCon 2021, the successfully held virtual meeting revealed a number of key lessons in digital meeting planning, international collaboration, and interdisciplinarity in astrobiology. Here, we report on the organization, structure, statistics, and execution of AbGradCon 2021 as a means to provide a successful case study to aid in the design of future virtual meetings, especially in astrobiology. As astrobiology is an inherently interdisciplinary field, AbGradCon 2021 introduced a unique interdisciplinary abstract evaluation procedure to determine the meeting speakers and participants; we report on statistics related to evaluation and understanding of interdisciplinary scientific abstracts gleaned from the abstract evaluation data. Finally, we close with a prospective of what aspects of the meeting we would recommend to improve organization, design, and execution of future meetings (Moss *et al.*, 2020, 2021).

## 2. Meeting Organization, Format, and Execution

### *2.1 Meeting Schedule*

As the major theme of AbGradCon 2021 was international collaboration, a goal of the meeting was to facilitate connection building between attendees from different countries. The conference schedule was thus divided into three major regions based on time zone (an oversimplification driven by necessity to fit within the Japanese time zone of the organizers (Japan Standard Time, JST)): East Asia/Oceania, Middle East/Europe, and The Americas. Each scientific session was scheduled to include two of these regions, so as to not marginalize attendees from specific time zones (which is an issue of non-inclusivity of many digital meetings (Jack and Glover, 2021)). Optimally, this meant that the conference would consist of three sessions spread out over the 24 hour day, but the decision was made to optimize for the organizers hosting and running the event who were based in Japan, resulting in two scientific sessions per day (**Tables 1–2**). One session would be in the morning JST, which allowed attendees from the regions of East Asia/Oceania and the Americas (and all areas in between) to attend at a "reasonable hour" ("reasonable hour" is loosely defined, spanning 0800 to 2230 in local time of the attendees). The second session would be in the late afternoon JST, allowing attendees from the East Asia/Oceania region and the Middle East/Europe region (and all areas in between) to attend at a "reasonable hour".

**Table 1. September 14 Session Schedule. The schedule of Sep 14 was slightly different from Sep 15-17, and as such the Sep 14 schedule is reported separately. The event times in JST, CEST (Central European Summer Time), and EST (Eastern Daylight Time) are reported to showcase the reasonability of the events for the selected regions' time zones.**

| Session | Time (JST) | Time (CEST) | Time (EDT) | Sep 14 |
|---|---|---|---|---|
| Morning Session | 0900-0920 | 0200-0220 | 2000-2020 (-1 day) | Introduction |
| | 0920-0950 | 0220-0250 | 2020-2050 (-1 day) | Flash talks |
| | 0950-1000 | 0250-0300 | 2050-2100 (-1 day) | Group Photo |
| | 1000-1030 | 0300-0330 | 2100-2130 (-1 day) | Break |
| | 1030-1130 | 0330-0430 | 2130-2230 (-1 day) | Keynote Presentation |
| | 1130-1600 | 0430-0900 | 2230 (-1 day)-0300 | Informal Interactions and Discussions |
| Afternoon Session | 1600-1620 | 0900-0920 | 0300-0320 | Introduction |
| | 1620-1650 | 0920-0950 | 0320-0350 | Flash talks |
| | 1650-1700 | 0950-1000 | 0350-0400 | Group Photo |
| | 1700-1730 | 1000-1030 | 0400-0430 | Break |
| | 1730-1830 | 1030-1130 | 0430-0530 | Keynote Presentation |
| | 1830- | 1130- | 0530- | Informal Interactions and Discussions |

**Table 2. Sep 15–17 Session Schedule. The session timetable for Sep 15-17 was identical for each day, although the content was different. Ample breaks were designed into the program for health reasons, so that participants could stand up from their terminals to stretch once in-awhile, as prolonged periods of remaining sedentary are detrimental to personal health** (Cowgill *et al.*, 2021)**. The event times in JST, CEST, and EST are reported to showcase the reasonability of the events for the selected regions' time zones.**

| Session | Time (JST) | Time (CEST) | Time (EDT) | Sep 15 | Sep 16 | Sep 17 |
|---|---|---|---|---|---|---|
| Morning Session | 0900-0945 | 0200-0245 | 2000-2045 (-1 day) | Live-streamed Presentations with Q&A | | |
| | 0945-0950 | 0245-0250 | 2045-2050 (-1 day) | Break | | |
| | 0950-1035 | 0250-0335 | 2050-2135 (-1 day) | Live-streamed Presentations with Q&A | | |
| | 1035-1040 | 0335-0340 | 2135-2140 (-1 day) | Break | | |
| | 1040-1110 | 0340-0410 | 2140-2210 (-1 day) | Small Group Discussions | | |
| | 1110-1130 | 0410-0430 | 2210-2230 (-1 day) | Report Discussion Results and Announcements | | |
| Break | 1130-1600 | 0430-0900 | 2230 (-1 day)-0300 | Informal Interactions and Discussions | | |
| Afternoon Session | 1600-1645 | 0900-0945 | 0300-0345 | Live-streamed Presentations with Q&A | | |
| | 1645-1650 | 0945-0950 | 0345-0350 | Break | | |
| | 1650-1735 | 0950-1035 | 0350-0435 | Live-streamed Presentations with Q&A | | |
| | 1735-1740 | 1035-1040 | 0435-0440 | Break | | |
| | 1740-1810 | 1040-1110 | 0440-0510 | Small Group Discussions | | |
| | 1810-1830 | 1110-1130 | 0510-0530 | Report Discussion Results and Announcements | | |
| Break | 1830- | 1130- | 0530- | Informal Interactions and Discussions | | |

Each plenary session (labeled "live-streamed presentations with Q&A") covered one topic through three presentations, and generally plenary sessions with the same topic appeared only two times throughout the course of the meeting (*i.e.*, once during a morning session and once during an afternoon session). Session topics varied widely, including biology, geology, astronomy & physics, planetary science, chemistry, and engineering & instrument design. This session topic allocation allowed the participants to be exposed to presentations from a variety of disciplines in astrobiology, which is inherently a very interdisciplinary field (Race *et al.*, 2012).

*2.2 Meeting Attendees*

260 abstracts from applicants based in 34 different countries were submitted for AbGradCon 2021, all of which were either undergraduate or graduate students, postdoctoral fellows, or other early career researchers. After a strict peer review process (*vide infra*), 146 of these abstracts were accepted in early June, and ultimately 119 attendees registered for the meeting, representing 26 countries of residence (30 countries of citizenship) (**Table 3**). The abstracts submitted by these attendees included diverse topics. However, the titles of all accepted abstracts included 29 mentions of "life", 17 mentions of "Earth", 16 mentions of "Mars", nine mentions of "origin" or "origins", and eight mentions of "RNA", which shows the areas which were more focused upon by the attendees; interestingly, "implication" or "implications" were also quite popular words, with nine mentions as well (**Fig. 2a**). Similarly, we analyzed the word-counts of all of the abstract texts, and found that they contained 252 mentions of "life", 232 mentions of words containing "chem" in any part, 150 mentions of "Earth", 138 mentions of words containing "environ" in any part, 105 mentions of "RNA", 100 mentions of words containing "evol" (short for evolution-related words) in any part, 90 mentions of "Mars", 89 mentions of "condition" or "conditions", and 80 mentions of "early"; "can" and "will" appeared 122 and 93 times, respectively (**Fig. 2b**).

**Table 3. Regions of residence of AbGradCon 2021 participants.**

| Region | Number of Participants |
|---|---|
| Asia | 28 |
| Europe | 22 |
| Middle East | 1 |
| North America | 61 |
| Oceania | 3 |
| South America | 4 |

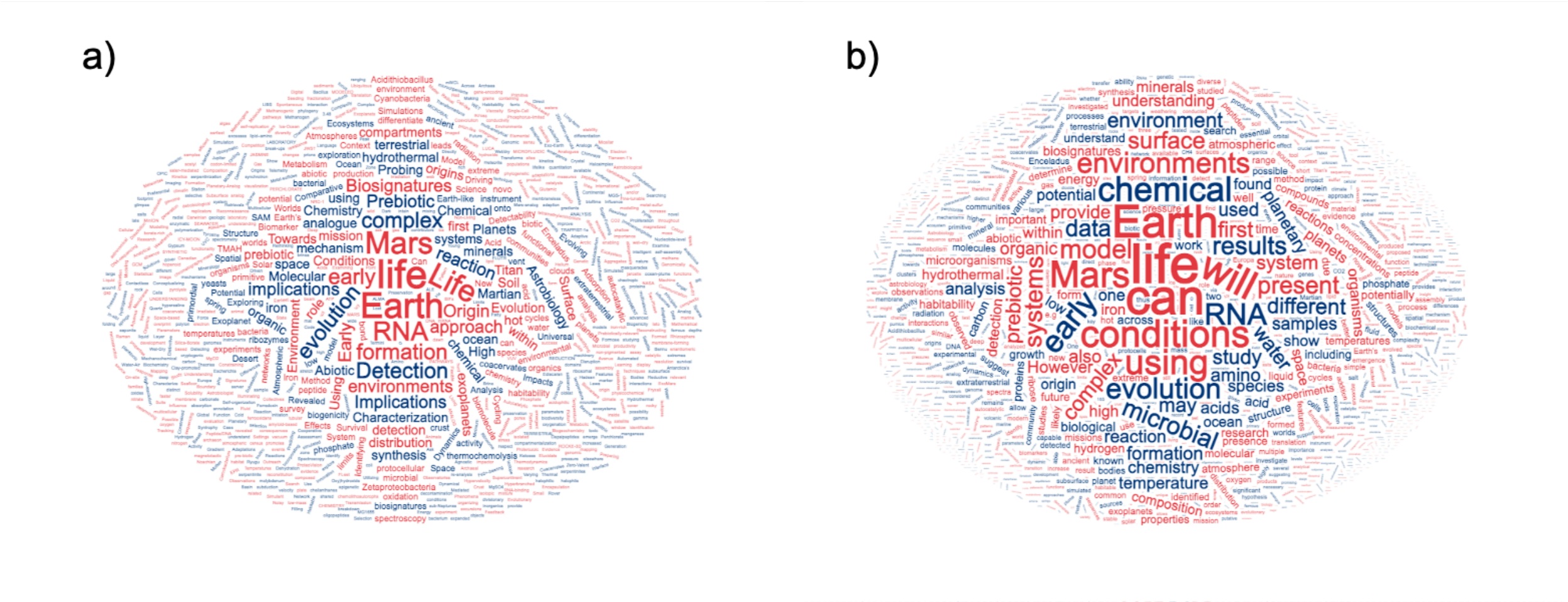

**Figure 2. Wordclouds (wordclouds.com) representing the frequency of words within the attendees' (a) abstract titles and (b) abstract text.**

For 96 of the 119 attendees, AbGradCon 2021 was their first attendance at an AbGradCon conference. 114 attendees reported gender/gender identity, including 56 women, 55 men, 1 nonbinary, 1 male/non-binary, and 1 agender participant. Each participant was also asked to self-disclose up to three fields that their abstract was relevant to in order to aid in sorting speakers and abstracts in the conference program. Abstract disciplines included such descriptions as astrobiology (which was then altered to the secondary choice of best fit), astronomy, biology, chemistry, education/outreach/diversity, engineering and instrument design, geology, physics, planetary science, and others (**Table 4**). Additionally, all participants were required to agree to a code of conduct, which required them to pledge to behave in a professional manner, to report observed harassment and misconduct, to protect privacy of prospective participants and their research (by not taking screenshots of presentations and not sharing information with those not attending the meeting).

**Table 4. Fields of abstracts presented at AbGradCon 2021, as self-reported by the presenters. Each abstract was required to be associated with at least a primary field, but association with a secondary and/or a tertiary field was not required (10 abstracts were not assigned a secondary field, while 45 abstracts were not assigned a tertiary field).**

| Field | Number of Abstracts (Primary Field) | Number of Abstracts (Secondary Field) | Number of Abstracts (Tertiary Field) |
|---|---|---|---|
| Astronomy | 7 | 4 | 3 |
| Biology | 42 | 19 | 10 |
| Chemistry | 27 | 26 | 13 |
| Education/Outreach Diversity | 2 | 2 | 1 |
| Engineering and Instrument Design | 4 | 0 | 3 |
| Geology | 10 | 17 | 7 |
| Physics | 3 | 5 | 8 |
| Planetary Science | 22 | 16 | 12 |
| Others | 0 | 20 | 16 |

*2.3 Pre-Recorded Presentations*

*In lieu* of completely live presentations, which have considerable risk of coming upon technical errors of all sorts, all participants of AbGradCon 2021 submitted pre-recorded videos as their presentations. Participants could choose to submit up to two types of videos: a 12-minute full presentation, or a 1-minute flash talk. All flash talks were streamed during a plenary session of the meeting, while a number of full presentations were selected for live-streaming during a plenary session with a short question and answer session immediately following the video. By selecting pre-recorded videos and streaming these videos from a central account, no time was lost due to presenter transition; as a result, no session went over-time. Participant presentations were also uploaded to Youtube as unlisted videos and the links shared to all participants (and not made public) so that they could stream them on-demand. About one month after the conference, videos of presenters that agreed to have them publicly released were published on the NASA Astrobiology Youtube channel, while videos of presenters who elected to not have them publicly released were deleted. In total, 36 full presentations and 44 flash talks were live-streamed during the conference, with an additional 61 full presentations available for on-demand streaming.

*2.4 Keynote Presentations*

One keynote speaker for each session (total of two speakers) on September 14, 2021 was invited to give a presentation entailing both their current research as well as their career path in astrobiology. In particular, as the meeting theme is international collaboration, desired keynote speakers would have extensive experience both working in different countries and collaborating with researchers around the world. Two speakers who satisfied these criteria (while also being based in Asia, which could further potentially connect attendees from outside of the Asian region with astrobiology research communities within the region) were selected.

Prof. Hikaru Yabuta of Hiroshima University in Hiroshima, Japan is an expert in cosmochemistry and the leader of the organic macromolecule sub-team of the initial analysis in JAXA's Hayabusa2 asteroid sample return mission (Watanabe *et al.*, 2019; Tachibana *et al.*, 2022). Her previous work experience has included appointments around the world.

Prof. André Antunes, who leads the astrobiology group at the State Key Laboratory of Lunar and Planetary Sciences at Macau University of Science and Technology (MUST) in Macau, China, is a microbiologist whose expertise is in astrobiology and exploring various extreme environments around the world, especially deep-sea brines (Antunes *et al.*, 2015; Antunes, Olsson-Francis and McGenity, 2020), and has had experience working in Europe, Africa, Asia, and the Middle East.

### *2.5 Breakout Discussions*

Live streamed presentations were followed by 30-minute breakout discussions at the end of each session. Participants were assigned to discussion groups randomly to allow for the representation of various research fields and countries in each group and were then invited to join the designated Gather.town (*vide infra*) discussion group spaces to meet with their peers. Groups always consisted of a discussion leader (a volunteer from within the participants or organizers), one of the speakers whose presentation was live streamed in the preceding session, and other meeting participants. A total of at most six discussion groups were running in parallel, ranging from three to ten participants each.

Discussion leaders were provided with prompting questions related to the topic and talks of the session and were encouraged to use these as a basis for further dialogue. Ideas and insights addressed during the discussions were collected and uploaded to the respective discord channels (*vide infra*), allowing participants from other groups as well as other time zones to read through them. Breakout discussions enabled participants to meet with one another and tackle specific and open questions through an interdisciplinary approach and were also used to reflect on the topics of each session. Key points were presented in a final overall discussion with all participants before the sessions were closed.

### *2.6 Gather.town platform in lieu of Zoom*

Gather.town is an online meeting platform which connects video conferencing with (virtual) spatial positioning of participants. This is done by placing a controllable avatar for each user in a shared virtual space. Gather.town attempts to replicate real-life interactions by making users move their avatars throughout the space and initiate video/audio connections based on the proximity of their avatar to those of others, or based on designated rooms with conferencing, and creates a more natural environment for gatherings compared to conference software like Zoom by utilizing shared virtual space. For example, rooms in Gather.town can have locations which are designated as spotlight, and users standing within a spotlight will stream their video or screen sharing to the entire room. In that way, the conference chair can very naturally say: “Can speaker Y come to the stage?”, while on classic conferencing software (Zoom, etc.) the chair must give permission for screen sharing to speakers. The same is true for Q&A sessions with spotlight locations, acting as real-life microphones, where Gather.town users can line up and ask their questions.

The platform provides several premade spaces and the ability to design your own custom space. This provides enormous flexibility in planning a conference by allowing organizers to connect and optimize planned activities with the space they will be executed in. As an online platform, Gather.town also provides capabilities which do not necessarily exist in real life, such as portals to reach different parts of the conference space quickly, locating users by their name and following them, private bubbles for communicating without interrupting the main presentation, and others. Gather.town has thus been implemented in various online fora up until now including education (Latulipe, 2021; McClure and Williams, 2021), workshops (Samiei *et al.*, 2020), and conferences (Jacobs and Lindley, 2021). However, at the time of first implementation (mid-2021), it was unclear whether Gather.town was GDPR (General Data Protection Regulation) compliant, an important regulation related to data storage and privacy for European Union (EU) or European Economic Area (EEA) residents and citizens as Gather.town is not able to save data in the EU. Thus, the organizers asked all participants to acknowledge this fact, and to sign a data processing agreement with Gather.town directly before registering for the meeting if this was a point of concern. As of January, 2022, Gather.town is now both GDPR and CCPA (California Consumer Privacy Act)-compliant. In addition, EU or EEA citizens and residents may directly make requests related to personal information or data storage, as well as make complaints or claims to local authorities.

For the purposes of AbGradCon 2021, a custom Gather.town virtual meeting space was designed, using the original meeting proposed space at ELSI as an inspiration for several common rooms (**Fig. 3**). The virtual meeting space included a plenary room, several breakout rooms for small group discussions, a media room with links to submitted presentation videos, an entertainment room containing games and social areas, and a main gathering room with multiple private spaces intended for informal discussions initiated by the participants.

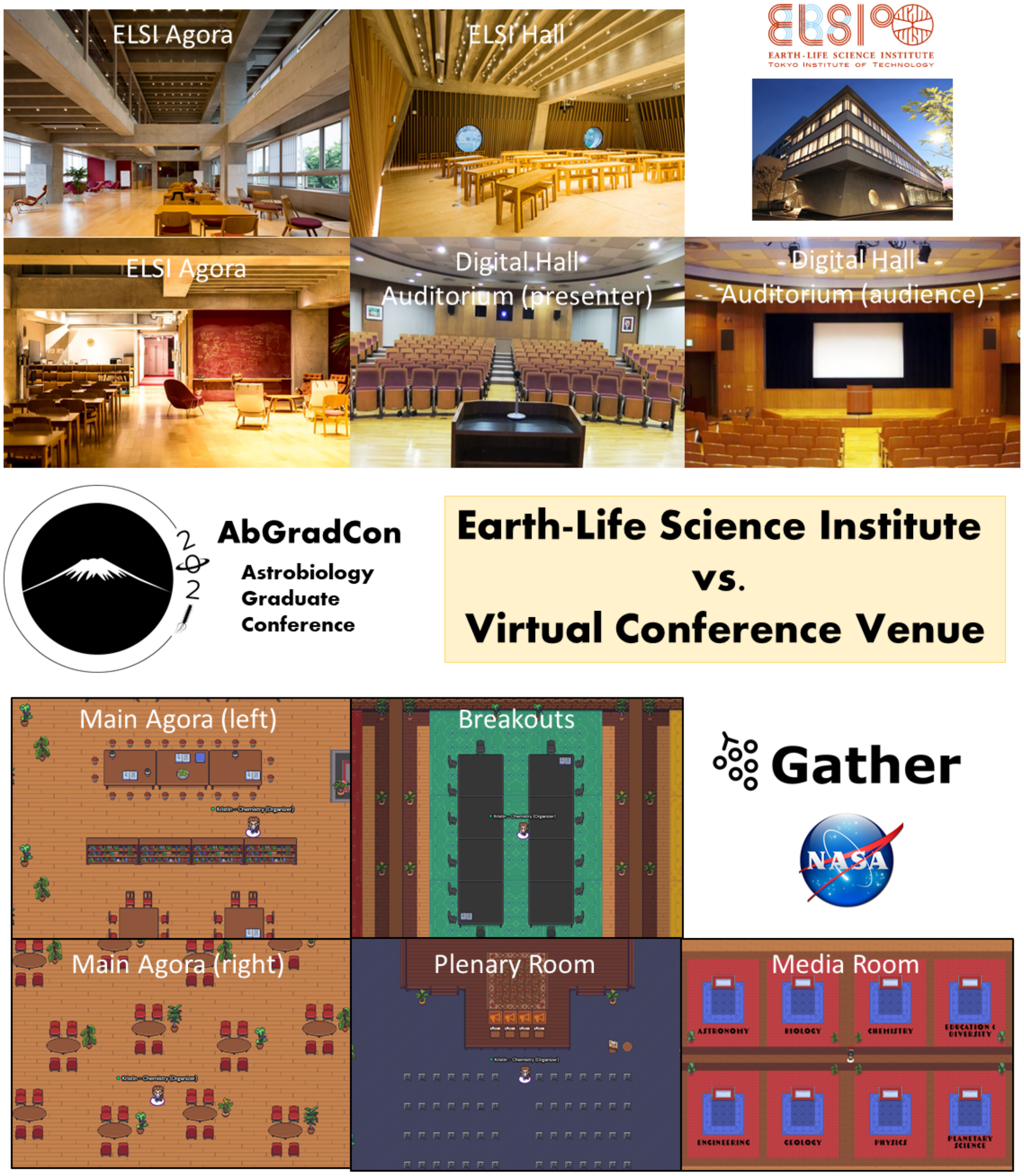

**Figure 3. Comparison of Virtual Meeting Space in Gather.town vs. The Original Proposed Meeting Space at ELSI, Tokyo Institute of Technology.**

*2.7 Discord Server*

To facilitate text-based discussions before, during, after, and between sessions, a Discord server was created as a mechanism to supplement the Gather.town platform. First of all, the Discord server was used by the organizers to make official conference announcements, to list the code of conduct, to list all other relevant links (such as to videos, the conference schedule, the Gather.town

platform, etc.), and also contained a help channel for participants with technical difficulties. A number of channels were also created for deposition of notes taken during the breakout discussion sessions (*vide infra*), while some private channels were made for exclusive discussion amongst the organizers regarding organizational issues. A number of public channels, which all participants could join and comment freely in, were also created covering a variety of topics. These topics included science communication, the proposal review panel (*vide infra*), a channel for participant introductions, career opportunities, upcoming meetings, general science, and a social channel. Participants were also able to propose new channels, and a number of channels dedicated to discussion on specific scientific topics were created for this purpose. Finally, Discord is both GDPR and CCPA-compliant.

### *2.8 Pre-conference Workshops and Activities*

As AbGradCon is a conference for students and early career researchers, career building is normally incorporated into the meeting agenda. A proposal writing retreat (PWR), for participants to practice preparing a scientific proposal, is usually held before the meeting. However, due to time zone differences and unbalanced availability of participants and organizers, such a PWR was not held before AbGradCon 2021. Instead, a number of alternative pre-conference workshops and activities to highlight important skills for researchers were held instead.

#### *2.8.1 Proposal Review Panel*

In the days before a traditional in-person AbGradCon, a PWR is usually held, gathering interested participants in an intensive multi-day workshop intended to introduce the NASA solicitation and proposal writing process. Writing is done in 3–4 person multidisciplinary groups, and groups are tasked with submitting a proposal abstract preceding the in-person event. When participants arrive, experienced organizers formally introduce them to the proposal writing and review process, after which groups embark on an intensive 48–72 hour professional development and writing exercise. By the end of the PWR, participants have written a complete NSPIRES (NASA's solicitation system) compliant proposal.

Due to restrictions on in-person activities as a result of COVID-19, rather than complete removal of the PWR component from AbGradCon, a remote PWR-replacement activity was held. This activity shifted the focus from proposal writing to proposal evaluation to ensure that the activity remained as practically useful as possible. Previously written and reviewed PWR proposals were selected and evaluated in groups (with various components redacted for privacy and after receiving the consent of their authors), led by organizers. The evaluations mimicked those of a genuine NASA review panel as closely as possible. Prior to these 2–3 hour group evaluations, participants took part in a 1 hour overview of the review process, led by an organizer with extensive NASA review panel experience. The proposals served as focal points of discussion between pre-selected groups as a mock review component and as a vehicle through which they can see both how typical proposals may be written, and how constructive feedback could be provided. This also included a focus on the budgeting and data management components of a proposal, to which students are not typically exposed or formally trained in, until faced with a real proposal. This allowed participants to understand the factors that can contribute to different costs and considerations throughout the lifetime of a project, including a focus on how time is budgeted among team members and institutions.

Despite condensing the typical 48–72 hours of in-person PWR activities down to only 3–4 hours of remote activities, participants were able to gain practical experience with an exceptionally realistic NASA proposal review process. Such experience is useful for scientists of any career stage, but especially relevant to AbGradCon's early-career scientists who are less familiar with the process, yet just as reliant on writing successful proposals for receiving funding. Feedback from participants highlighted how valuable the information and experience was. Participants went so far as to say they were surprised that such information was not previously supplied during their graduate careers. By continuing to include PWR or a proposal-related activity in future iterations of AbGradCon, and with the ability to quickly shift to remote participation if need be, this unique training opportunity can be made more widely available which can increase the quality of highly-relevant but oft-taught early-career science education across the diverse and multidisciplinary astrobiology community.

*2.8.2 Proposal Budgeting Exercise*

Proposal budgeting is often an important task in the proposal preparation process, yet many early career researchers have no experience preparing budgets by the time they must start their first proposal. In addition, formal education during programs in graduate or postdoctoral training often focus solely on research rather than important administrative and bureaucratic skills that are required for long-term success as a researcher. As such, a proposal budgeting exercise was prepared for participants to get a taste of what preparing a proposal budget is like.

Three practice proposals prepared during a previous PWR (2019) were selected (in three different fields). Similar to the proposal review panel, all personal identifiable information was deleted from all proposals. Additionally, while the line items in the prepared budgets were retained, all costs were deleted, and these modified proposals were sent to participants, who were given three weeks to read all proposals and fill in the costs of the budget line items. Later, the actual costs from the proposals were given to the participants for their records so that they could evaluate their own budget cost determination skills.

*2.8.3 Science Communication Workshops*

Acquiring science communication skills can be quite beneficial throughout one's research career as a student and a scientist. However, workshops and training catered towards building science communication skills for students and researchers are still limited. Recognizing this need, the organizing committee used AbGradCon 2021 as an opportunity to provide additional science communication training, in the form of workshops and recorded video sessions, to the participants.

The committee collaborated with Explainables (explainables.org), professional science communication trainers, to develop workshops under two themes: 1) Authentic Branding, Professional Networking and Marketing for Scientists and 2) Multidisciplinary and Cultural Communication 101. The workshops were held prior to the main conference in order to prepare the participants for the conference, and organized for small groups (up to 15) at different times throughout the day so that participants from different time zones could be able to join. Organizers also took into account fatigue from attendance of long online meetings and conferences, and thus the workshops were developed as virtual hands-on experiences to create an active engagement, while having a maximum time-limit of four hours and including ample break times. Moreover, Explainables trainers provided recorded videos covering two themes: 1) Best Practice for Creating Scientific Presentations and Posters, and 2) Branding, Networking and Marketing for Scientists.

These videos were accessible to all participants before, during, and after the meeting, while public viewing was possible for about two months after the meeting finished.

*2.9 Social Media*

During the organization phase of both the 2020 and 2021 meetings, the Twitter and Facebook accounts of AbGradCon were used for various purposes, such as raising interest about the conference, announcing deadlines (*e.g.*, abstract submission), updating (prospective) participants and the community on decisions taken following the COVID-19 pandemic situation, as well as sharing job opportunities, conferences, and recent astrobiology papers. For example, the AbGradCon social media accounts were used to share the experiences of former attendees and organizers of past AbGradCon conferences. This was done as a way to motivate early-career astrobiologists to attend AbGradCon 2020/2021 by showing how taking part in the conference has helped past attendees to get in touch with other researchers in the field and to pursue careers, both within and outside of academia. Furthermore, a number of organizing team members were also featured on social media before the conference with a photo and a short description about their motivation to take part in the organization of (and/or past participation in) the conference to highlight their contribution to organizational efforts. Finally, social media was also used during the conference as a way to share information and updates about the conference (*e.g.*, upcoming sessions, conference photos on Gather.town) with the participants and the wider community. By using the hashtag #AbGradCon21, participants highlighted the work presented at the virtual conference and shared their experience.

*2.10 Conference Booklet*

A conference booklet was prepared to guide the participants through the meeting as a supplemental resource. This booklet contained basic information about the conference organizers, host (ELSI), and sponsors/partners. It also provided links to a variety of supplementary resources including science communication video workshop sessions produced by a partner organization, the links to the various platforms (Gather.town and Discord) along with tutorial videos/resources for both platforms, and links to video recordings (Youtube) from previous AbGradCon conferences. In addition to the program, the abstract and biography of the two keynote speakers was provided.

The conference booklet also acted as an abstract booklet. First, all participants, their institutions, and their presentation title were listed by discipline for ease of searching by discipline. Then, grouped by discipline, each abstract was listed including the contact information of the presenter (if they chose to have this disclosed to other participants) and the link to the presentation and/or flash talk (Youtube). In this way, participants could easily access videos or contact the presenter immediately after reading the abstract. Finally, a simple table of all participants in alphabetical order (by last name) was compiled as well as any video links associated with them was presented as a means for accessing video links by presenter rather than abstract.

### *2.11 Conference Gift Package*

In many conferences, a conference gift bag is distributed to each participant containing conference information and also items from and information about the sponsors. AbGradCon 2021 elected to send a small gift package before the meeting to all participants who would accept one as a means to increase anticipation before the meeting. A number of sponsors contributed gift items to the gift package, and all of the items were sent to a centralized location in the United States. Then, these items were arranged and distributed into separate packages for further distribution by domestic or international shipping to the relevant participants (with support from AbGradCon partners Breakthrough Listen, Blue Marble Space Institute of Science (BMSIS), and Greenspace). Ultimately, while many of the gift packages were delivered before the meeting, due to the current situation regarding the tardiness of international shipping, a number of gift packages were not delivered until after the meeting began or concluded.

### *2.12 Sponsors and Partners*

AbGradCon 2021 could not have been successfully held without the support of various sponsors and partners (listed in detail in the acknowledgements). Some of these sponsors and partners provided funding for the Gather.town platform, hiring workshop facilitators, and payment for conference gift package shipping, while other partners managed and distributed this funding (a number of sponsors pledged funding for both the 2020 (postponed) and 2021 meetings, but ultimately this additional funding was not needed). Other partners supported the meeting through advertisement of the meeting to their networks, leading and facilitating workshops, and packaging and sending the conference gift packages.

*2.13 2020 Undergraduate Flash Talk Contest*

In 2020, *in lieu* of the traditionally-held Undergraduate Poster Contest (which grants travel funding to the conference to the winner), the organizers decided to hold an Undergraduate Flash Talk Contest. Undergraduate students contributed a short flash talk (audio only) describing their research in five minutes or less. Each of these flash talks was evaluated based on content (clarity, quality, and how well their content supported their claims), organization (transitions, logical flow, clear thesis and supporting data, and informative and clear project summary), and delivery (professional/engaging delivery, accessibility to those outside of their field, and enthusiasm). Ultimately, two co-winners and one honorable mention were selected, and these flash talks were featured on the AbGradCon website. Originally, only the co-winners would receive full travel funding to attend AbGradCon 2020, but given its postponement and the digital nature of AbGradCon 2021, all co-winners and the honorable mention were granted acceptance to AbGradCon 2021 if they chose to submit an abstract.

*2.14 Organizational Roles*

The tasks for the organization of both the 2020 and 2021 meetings were shared among different members of the local and external organizers; each role provided support for various essential aspects for the meeting. Some roles were filled by one organizer, while others were filled by a team of organizers; all organizers also contributed to multiple roles. Here, we provide a brief overview of each of these tasks listed in somewhat chronological order, separated by meeting year, as well as the rough number of organizers that contributed to that role (in parentheses after each role). We note that while some organizers selected roles which allowed them to learn a new skill, ultimately most of the organizers performed roles which took advantage of their already-held skills and experience. After the postponement of the meeting from 2020 to 2021, many organizers needed to step down from the event, resulting in more labor falling to a smaller group of organizers. As a result, many of the roles which began as well-organized group experiences changed to the responsibility of one or two organizers taking on multiple roles, unable to easily recruit new team members to replace those who stepped down. Luckily, many background aspects of the organization from 2020 (*i.e.*, logo, webpage, sponsors and partners, application, abstract evaluation scheme, etc.) were able to be carried over to 2021 without significant variation (although for certain aspects, like the application, lessons from the 2020 meeting application process were used to optimize the 2021 application), while other components (*i.e.*, field excursion and undergraduate flash talk contest) were abandoned by necessity. Additionally, roles were given to professional partners to develop new interactive components for the digital meeting.

*2.14.1 Roles Present for Both AbGradCon 2020 and 2021*

**Proposal Preparation (2 organizers).** A funding proposal to hold AbGradCon and the PWR was prepared and submitted to the NASA Astrobiology Program to cover travel, accommodation, and other costs for participants.

**Logo design (3 organizers).** A logo for the 2020 and 2021 meetings (**Fig. 1**) was designed in collaboration with a graphic designer (Lucy Kwok, http://be.net/thislucykwok). The AbGradCon logos incorporated both aspects related to astrobiology and the physical location of the conference (Japan).

**Soliciting Funding and Sponsors (2 organizers).** In addition to the generous funding provided from the NASA Astrobiology Program, additional funding from non-governmental organizations, such as journals, scientific societies, and other private organizations were solicited to support the meeting through funding, often with less restrictions (often for use for international travel).

**Communication with Partners and Sponsors (2 organizers).** Partners and sponsors had regular communication with the organizing team for funding updates, as well as for communication about any official conference updates.

**Budget Management (1 organizer).** The AbGradCon budget was managed in concert with BMSIS (for US-based funding) and ELSI (for Japan-based funding).

**Website design and management (4 organizers).** A conference website was designed using the Google Sites platform (migrated from the previous GitHub platform), as a way to provide some background about astrobiology and to communicate announcements, updates (especially relative to the COVID-19 pandemic), and other relevant information about the meeting (activities, host institute, sponsors, code of conduct, past meeting websites, etc.). This website was updated regularly with the latest information not only for the participants, but also for the public.

**Social Media (4 organizers).** As mentioned previously, social media was used to make official announcements and to disseminate job, research, and funding opportunities both from our partners/sponsors as well as from the astrobiology community at large.

**Advertisement (all organizers).** The meetings were advertised by email, on social media, on the conference website, as well as using posters and leaflets that were distributed at other conferences and workshops. Each organizer also distributed the conference information to their relevant networks including institutes of study/employment, research collaborators, and scientific societies.

**Abstract Submission Organization (2 organizers).** The abstract submission process, through Microsoft Forms (due to Google connectivity issues in China), was designed and included various categories of applicant demographics information for use in future meetings. The submitted abstracts were organized by members of the team until abstract evaluation.

**Abstract Evaluation Schema Design (4 organizers).** The parameters by which the abstracts were to be evaluated (*vide infra*) were decided, including how much weight would be put into each category.

**Abstract Evaluation Lead (2 organizers).** The abstracts were evaluated by external reviewers, which were recruited and led/managed by the organizing team. This included giving reviewers instructions, distributing the abstracts to be reviewed, and compiling the review reports of the reviewers. In the case some reviewers could not evaluate all abstracts, those were also reassigned to other reviewers.

**Participant and Speaker Selection (1 organizer).** After the abstract reviews were received, the evaluation data was organized; from this information, which applicants would be selected for live-streamed presentations, for attendance, or not accepted were decided. Finally, for AbGradCon 2020, for those who were accepted, travel funding was divided and planned to be disbursed to participants with the highest scoring abstracts (before cancellation).

**Invitation and Organization of Keynote Speakers (2 organizers).** Keynote speakers were selected and invited to give presentations during the conference. Communication and coordination with the keynote speakers included scheduling the presentations and instruction on how to use the online platform for their presentations.

**Conference Communication (2 organizers).** Official conference announcements to the applicants and participants (after selection) were communicated through email (and later, through the Discord server). This role also included providing support or answering questions for applicants/participants in case there were any questions or issues with the abstract submission process, the registration process, or any aspects related to the conference itself (such as providing an official acceptance letter).

***2.14.2*** *Roles Present only for AbGradCon 2020*

**Field Trip Coordination (4 organizers).** We had planned for field trips to visit three local research centers specializing in different aspects of astrobiology research, each with unique aspects which may have been interesting to the participants: National Astronomical Observatory of Japan (with a visit to the 4-dimensional digital universe (4d2u) projection) (Usuda-Sato, Tsuzuki and Yamaoka, 2018), Japan Agency for Marine-Earth Science and Technology (with a visit to a research vessel), and Institute of Space and Astronautical Science of JAXA (with a visit to the Communication Hall of Space Science and Exploration). These field trips required negotiation with representatives from each site, as well as coordination of travel plans and planning of activities. Because of the online nature of the meeting, no field trips were held during AbGradCon 2021.

**PWR Organization (6 organizers).** The original PWR was planned to be held at a resort about 1.5 hours away from the main conference site. This required organization not only regarding the logistics of accommodation and meals during the retreat (including negotiations with the resort) as well as travel to/from the retreat site, but also organization of the content of the retreat, including introductory presentations, team management and mentoring, and submitted proposal evaluation. Ultimately, the AbGradCon 2021 PWR was not held in the format of past years due to the online nature of the meeting.

**Accommodations and Venue Logistics (3 organizers).** Normally, the accommodation site is chosen to be on-site, either at a University-affiliated hotel or a University lodge/dormitory. Tokyo Tech does not have such an option, and thus negotiations with external hotels were performed to secure accommodation for participants at a location reasonably close to the conference venue (before cancellation). Finally, conference venue negotiations (for internal Tokyo Tech and ELSI facilities) were completed as well before cancellation.

**Meals Logistics (2 organizers).** Meals for the conference were planned to be organized through a mixture of catering (from the University and local restaurants) and self-led exploration of local restaurants. Specific care was taken to satisfy food restrictions of attendees, as food restrictions in Japan are neither very common nor well understood by the general public or restaurants (as such, accommodations are limited). For example, pure vegetarian food is not easy to find in Japan (Freedman, 2016), although choices have increased in recent years.

**Undergraduate Flash Talk Contest (4 organizers).** Organizers designed the Undergraduate Flash Talk contest held in 2020, including the contest parameters and evaluation criteria. Then, the submissions were acquired, evaluated, and the final scores were tabulated to determine the winner(s).

**Solicitation of AbGradCon "Ask an Astrobiologist" panel (2 organizers).** After cancellation of AbGradCon 2020, the NASA Ask an Astrobiologist program (Impey, 2021), a monthly online video program featuring questions and answers to a panelist or panelists in the astrobiology field, organized an AbGradCon edition of their program. Previous organizers from around the world were solicited to join this program in order to discuss the internationality of astrobiology and to help to maintain interest in AbGradCon 2021.

### ***2.14.3*** *Roles Present only for AbGradCon 2021*

**Proposal Review Panel Organization and Execution (7 organizers).** The proposal review panel was conceptualized and designed accounting for the missing PWR, which was not held in 2021. Proposals from previous PWRs were curated and distributed to participants following an introductory session from an invited expert. Afterwards, panel leaders helped to guide participants through a mock review panel that evaluated the proposals, while the invited expert was on hand to answer any specific questions from participants (and panel leaders).

**Proposal Budgeting Exercise Organization (1 organizer).** The budgeting exercise involved curating proposals from previous PWRs, deleting personal identification information and budget information, and distributing this to the participants, followed by distribution of the proposal with budget information included for participants to self-check their results.

**Science Communication Content Organization (3 organizers and 3 members from a partner organization).** In collaboration with Explainables, science communication workshops and videos were organized by first deciding the content to be covered and the format that the content was to be delivered in (*i.e.*, workshops or videos). Then, collaborative work with the workshop facilitators included official announcements to participants as well as providing the facilitators with participant contact information for scheduling and communication.

**Curation and Distribution of Conference Gift Bag (2 organizers and various members from partner organizations).** In collaboration with various partners, gift bag items were curated (from partners/sponsors) and sent to a central location. Then, the addresses of the participants who requested a gift bag were organized and distributed to Greenspace, who was in charge of distribution of the gift bag internationally *via* post.

**Design and Management of Gather.town Space (3 organizers and 1 member from a partner organization).** The Gather.town space was designed (from scratch) by a NASA-affiliated staff member, with some input from some of the organizers. Then, the organizers were trained in proper usage of the conference platform so that they could troubleshoot before and during the conference.

**Design and management of Discord server (3 organizers).** Organizers conceptualized and designed the Discord server, including the initial channel categories to be included, and constant management was performed immediately before and during the meeting. This included answering questions related to troubleshooting, making official announcements, and also creating new channels when requested.

**Solicitation, Organization, and Uploading of Participant Videos (2 organizers and 1 member from a partner organization).** Pre-recorded presentation videos were solicited from the participants, which included communication into the format and parameters of acceptable videos. After the videos were all submitted by the participants, the videos were organized by category and uploaded to Youtube by a NASA-affiliated staff member. Finally, the Youtube links were collated and distributed to the participants before the meeting; after the meeting, the videos of participants who elected to publicly publish them were published, while the other videos were deleted.

**Session chairs (6 organizers and 1 participant).** Session chairs were solicited amongst the organizing team and other conference participants. The session chairs were then given the information about the speakers of the session, and were instructed to introduce each speaker. Because of the unique digital format of the meeting, all live-streamed presentations were streamed to all participants from a central location (one organizer's role was to deliver this content). As such, after introducing each speaker, the session chair needed only to curate the question and answer session and make sure that time was kept properly (including breaks). Finally, the session chair gave brief closing remarks at the end of each session.

**Discussion Organization and Leading (5 organizers and more than 20 participants).** Discussion leader volunteers were first solicited in the following order: organizational team members, previous AbGradCon attendees, senior attendees of AbGradCon 2021, and finally all attendees. Then, each discussion leader was assigned a session based on their availability, and leaders were given information about their discussion session, including good practices into how to lead discussion and encourage constructive dialogue, the day/time and location of their session, and prompting questions based on that day's talks. During the discussion, the discussion leaders were instructed to select a note-taker, who would catalog the topics in that day's discussion for deposition in the Discord server, as well as a presenter, who would give a short overview to all participants immediately after the discussion session. All participants were assigned into discussion sessions arbitrarily and informed immediately before the discussion session.

**Selection of 2022 Meeting Site (1 organizer).** First, bids were solicited from potential sites, and asked to provide a short presentation highlighting their vision for an AbGradCon organized at their site. These presentations were distributed to organizers and abstract evaluators, followed by voting by the organizers and abstract evaluators. After voting, the results were aggregated, and the winning site was announced. For AbGradCon 2022, only one site bid (Washington, DC, USA), and thus rather than voting between different sites, voters instead voted “Yes” or “No” for this site to host AbGradCon 2022. Important information about the funding, logistics, and organization of the AbGradCon 2021 were then transferred to the AbGradCon 2022 organizers.

**Post-Conference Debriefs and Reports (a majority of the organizers).** Following the conference, debrief sessions or reports to the various partners and sponsors were prepared. These included presentations, discussions, news articles, and other reports (such as the current one).

## 3. Abstract Evaluation

### *3.1 Abstract Evaluation Flow*

AbGradCon is a conference designed for a small community to participate in direct discussions and interactions with other participants. As such, traditionally, the total number of attendees has been limited (to around 100). Given the desire to contribute a similar conference feeling, AbGradCon 2021 would also cap the total attendance (also in order to balance the geographic distribution of attendees). Additionally, live-streamed presentations must be selected from submitted abstracts with some discernable metric. As such, an abstract evaluation process was used to assign a score to each submitted abstract; abstracts below a certain score would not be selected for the meeting, while the abstracts above this score would be accepted. Additionally, taking into consideration the fields of the abstracts, the highest scoring abstracts in a given field were selected for live-streamed presentations.

A number of potential volunteer abstract evaluators were contacted to contribute to this organizational step. These evaluators may have been past AbGradcon attendees, AbGradCon2021 organizers, members of partners of the current meeting, and other students/postdocs/early career researchers within the astrobiology community; evaluators were not chosen in any scientific or systematic way.

Each volunteer evaluator was initially assigned to between 20–21 abstracts arbitrarily and not delineated by field (all personally identifiable information, including author names and affiliations, were not included to maintain strict double-blind reviewing; we provided only the title and the abstract, strictly truncated to 250 words). If an evaluator noticed any conflict of interest based only on the title and the abstract, such as the abstract being their own, of a recent collaborator, of a lab member, or any other reason (such as a competing research group), they were asked to declare their conflict of interest and were assigned a supplementary abstract to evaluate by the organizers. Then, all evaluators were to proceed to evaluate each abstract according to a standard scoring rubric (*vide infra*), and report these scores in a central system. In total, each abstract was to be evaluated by three or four evaluators. However, there were cases where evaluators did not do any evaluations, had to drop out from this duty due to other reasons, or evaluated the wrong abstracts (those not explicitly assigned to them), which necessitated additional reviewing of a certain number of abstracts by evaluators; as such, the organizers assigned these additional abstracts with missing evaluations to evaluators who could volunteer to evaluate more abstracts. Finally, the average scores of all abstracts were tabulated, and conference acceptance and live-streamed presentation selection was determined based on the average score, field represented, and country/region represented (*vide infra*). All organizers, discussion leaders, and abstract evaluators were accepted to the meeting. A similar process was utilized for evaluation of abstracts for AbGradCon 2020, although the selection step after collation of all evaluation scores was halted due to the postponement of the meeting at that time.

### *3.2 Abstract Evaluation Scoring Rubric*

We included the following questions on the scoring rubric for abstract evaluations for AbGradCon 2021 (a similar scoring rubric was used for AbGradCon 2020 evaluations), with a brief explanation below.

**Conflict of Interest: Do you have a conflict of interest regarding this application? Examples: Do you share current or recent (last 2 years) research collaboration. Are you family with the applicant? Do you have a financial interest in the applicant? Did you help the applicant with their abstract submission?**

- Yes
- No

This question was included so that applicants could declare if they had a conflict of interest with any of their assigned abstracts so as not to bias the scoring.

**Is this work relevant to Astrobiology?**

- Yes
- No
- Maybe

Although AbGradCon aims to be an open conference for those who are interested in learning astrobiology for the first time, one of the abstract submission requirements was to include something that was astrobiology-related. For those who had abstracts in fairly far-afield topics, this could be achieved by proposing how one could use one's own research and apply it to astrobiology. While this question was not scored explicitly, any abstracts with all or nearly all evaluators selecting "No" to this answer were not considered for the meeting.

**What is the scientific field of this abstract?**
**Is there another field that this abstract falls under?**

As all evaluators self-reported their own primary and secondary fields, these two questions were included so that we could analyze any trends or biases that arise from evaluation of abstracts within vs. outside of one's own field of expertise. The fields that could be chosen are represented in **Table 4**, with an "other" option also available.

**Scientific Merit of the Abstract (Rate from 0-4).**

**Scientific Literacy (Rate from 0-4).**

**Novelty of the Theory/Work (Rate 0-4).**

**Soundness of the Theory/Work (Rate from 0-4)**

- **0- None**
- **1- Weak**
- **2- Average**
- **3- Strong**
- **4- Excellent**

These four questions are where the total scoring of the abstract were derived from (all four questions had identical choices). We designed these categories so that Scientific Merit rated the overall significance of the research, Scientific Literacy rated the background of the research, Novelty rated how novel the research was, and Soundness rated whether the experimental or theoretical methods were the correct ones to be applied to the research as well as the validity of the conclusions. However, evaluators were free to interpret these categories freely based on their own knowledge, understanding, and experience. During the evaluation process, a number of abstract evaluators expressed concern over their ability to evaluate these categories for abstracts outside of their own field. However, as astrobiology is inherently interdisciplinary, one of the criteria for submitted abstracts was to make it as understandable as possible for those outside of one's own field, and submitted abstracts which were more easily understood by a variety of evaluators from different fields likely scored higher.

**Does the applicant demonstrate competency in English?**

- **Yes**
- **No**
- **Other**

AbGradCon is held in English, but participants of all language backgrounds are welcome as long as they demonstrate the ability to competently communicate during the meeting in English. This question was not meant to discourage the participation of non-native English speakers. Rather, this question is meant to assess if the applicant's English level will result in difficulty communicating with other AbGradCon participants. If the evaluator was uncertain but had some comments, they could leave some notes.

**Do you have any other comments or concerns about this abstract? If yes, please describe your comments or concerns.**

- **Yes**
- **No**

This question was meant for evaluators to leave any further comments or notes about the abstract that did not fit any of the evaluation questions; these comments were used in case decisions had to be made about abstracts with similar average scores.

## 4. Revealing Data from Abstract Evaluation on Interdisciplinarity and Internationality of the Astrobiology Research Field

### *4.1 AbGradCon 2021 Submitted Abstract Information*

260 abstracts were submitted, and 225 of these abstracts were determined to be "valid"; abstracts were deemed invalid due to various reasons such as not providing a proper title and/or abstract, clearly not providing a professional abstract, not properly reporting country of institute or residence (this was a requirement for abstract submission, but was only used for statistical purposes), or submitting identical abstracts as another applicant, among others. Applications were submitted from participants residing in 36 countries (region breakdown is reported in **Figure 4a**) and with citizenship in 42 countries, while their primary fields (**Fig. 4b**) and career levels (**Fig. 4c**) were also broad, spanning from undergraduate student to early career independent researcher (*i.e.*, non-tenure track faculty). Finally, 140 of the applicants learned of AbGradCon from their colleagues, 22 from Twitter, and 30 from the AbGradCon website.

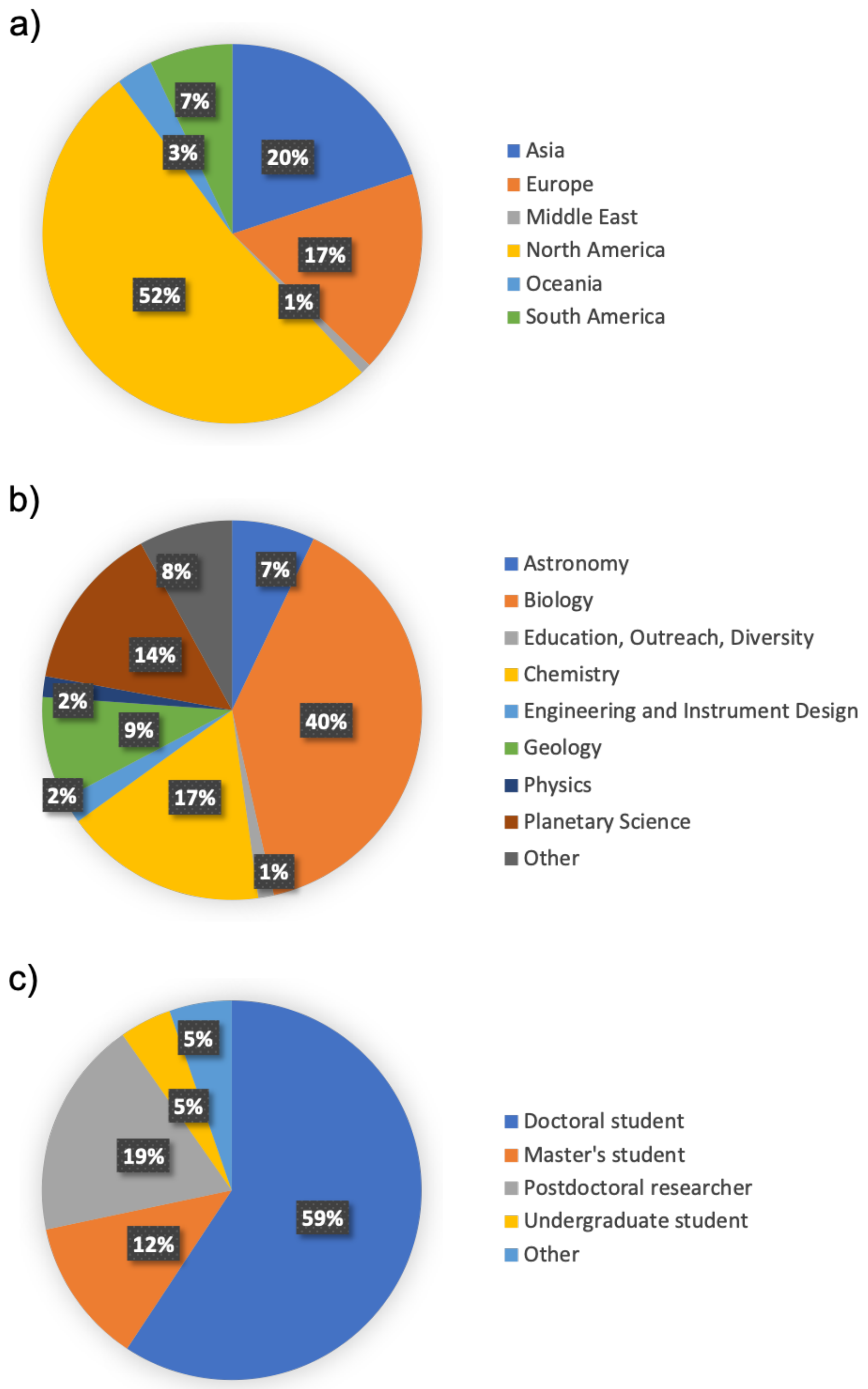

**Figure 4. Various demographic information about the applicants. a) Region of residence. b) Primary Field of submitted abstract (self-reported). c) Career level.**

### *4.2 Abstract Evaluator Information*

As mentioned previously, a number of evaluators, with a large breadth with respect to career level, region, discipline, and years of astrobiology experience, scored the abstracts. For example, evaluators ranged from master's students to postdoctoral fellows; were based in Asia, Europe, North America, and Oceania; reported primary fields ranging from astronomy, biology, chemistry, geology, and planetary science; and had been in the astrobiology field between 1 and 21 years. This suggested that the evaluation process could proceed fairly in principle, although no systematic mechanism for evaluator selection (such as evenly distributed amongst regions, fields, career levels, etc.) was performed.

After the evaluation process, the mean number of abstracts evaluated per evaluator was 25.5, with a standard deviation of 7.9. The maximum number of abstracts that any one evaluator evaluated was 42, while the minimum was 13. The evaluators had fairly different average evaluation scores. For example, one evaluator had an average evaluation score of 9.5 (out of 16 points, maximum), while another had an average evaluation score of 15.5 (the mean average evaluation score of all evaluators was 12.1, with a standard deviation of 1.8). Additionally, each evaluator had a different evaluation scoring variance, with some showing little variance in scoring (standard deviation as low as 0.7, or as high as 5.3). These observations of statistical variability between evaluators may be due to a number of different reasons, including interdisciplinarity of their own research (as abstracts evaluated may not have been from their own fields), however it is difficult to make definitive conclusions.

### *4.3 Abstract Evaluation Results*

Each of the 225 abstracts was evaluated at least three times by different evaluators, and on average by 4.1 different evaluators. However, three abstracts were evaluated by seven different evaluators, while 72 abstracts were evaluated by only three different evaluators. In total, there were 917 evaluations performed, with a global mean score (out of 16 points) and standard deviation of 12.1 and 3.1, respectively. 142 evaluations received a perfect score, three evaluations received a score of 0, and 12 evaluations were determined to be irrelevant to astrobiology. Evaluators were also asked to categorize each abstract that they evaluated into primary and secondary scientific fields (**Table 5**). It appeared that evaluations of abstracts determined to be in most of the "traditional" scientific categories (by primary field) were similar in score, but abstracts determined to be in the philosophy of science or society category had lower mean evaluation scores. A heteroscedastic two-tailed t-test comparing the scores of evaluations of philosophy of science abstracts against all other abstracts, and society abstracts against all other abstracts, showed that while the mean evaluation score of society abstracts was lower than that of all other abstracts (p-value of 0.031), the mean evaluation score of philosophy of science abstracts was not significantly lower than that of all other abstracts (p-value of 0.064).

In most conferences, abstracts are evaluated by experts in that particular field. Given that astrobiology is a broad field by definition with many sub-disciplines, abstract evaluators for AbGradCon2021 likely evaluated abstracts not only within their own discipline, but also within other disciplines where they were less familiar with. Thus, it may be possible that the lack of familiarity of topics of other disciplines led to systematic or random biases in evaluation of abstracts from out-of-discipline (especially for society-related abstracts), or perhaps even led to changes in evaluation score variation (due to potential “random” evaluation).

**Table 5. Statistics of Evaluations of Abstracts Sorted by Category (As Determined By Evaluators). The mean score of all evaluated abstracts was 12.1.**

| Primary Field | Mean Evaluation Score | Standard Deviation of Evaluation Scores | Number of Evaluations |
|---|---|---|---|
| Astronomy | 11.7 | 2.8 | 73 |
| Biology | 12.1 | 3.3 | 384 |
| Chemistry | 12.8 | 2.6 | 197 |
| Education, Outreach, and Diversity | 11.8 | 4.5 | 12 |
| Engineering and Instrument Design | 12.6 | 2.4 | 42 |
| Geology | 12.2 | 2.7 | 80 |
| Philosophy of Science | 8.6 | 4.5 | 8 |
| Physics | 10.9 | 2 | 17 |
| Planetary Science | 12.2 | 2.8 | 77 |
| Society | 7.4 | 4.4 | 7 |
| Other | 10.1 | 4.4 | 20 |

After all of the evaluations were completed, the results were compiled and the mean evaluation score of each submitted abstract was used to determine which abstracts would be accepted to the conference (*i.e.*, final abstract score). The maximum score for any submitted abstract was 15.67 (two abstracts), the minimum score was 5.2, and the mean and standard deviation of final abstract scores was 12.1 and 2.0, respectively. We next sorted each abstract based on various categorizations (region of applicant residence, primary field of abstract, or career level of applicant) to observe whether there were any trends related to evaluation of abstracts from different categories (**Table 6**). Abstracts from those in Asia on average appeared to receive lower evaluation scores than other abstracts, while abstracts from those in North American on average appeared to receive higher evaluation scores than others (with statistical significance). We speculate that this may be due to a number of cultural and language differences related to education practices in each country and the maturation of astrobiology specific scientific programs in those countries. For example, North America was one of the first regions around the world to adopt astrobiology as a research topic, and are currently world leaders in terms of astrobiology research productivity, funding, and education (with a number of astrobiology-focused programs). Although there has been expansion of astrobiology as a research and education topic in Asia, these activities have only gained significance in very recent years, and there are fewer astrobiology research groups, education programs, and funding opportunities in Asia as compared to North America or the rest of the world. As such, perhaps the quality of the abstracts is affected by the educational and research opportunities afforded by each student or researcher's home country/region; *i.e.*, without the proper training and opportunities, it is difficult to prepare a professional-level abstract for an international, interdisciplinary astrobiology conference. A more detailed study into those factors is beyond the scope of this report, but as the field of astrobiology matures and ages globally, such variations in abstract scores are likely to disappear.

Additionally, given that AbGradCon is traditionally organized in and by residents of North America, it may be that the abstract style required for submission or the abstract evaluation criteria are biased towards those who are either from or educated in North America. This regional scoring disparity may thus mean that the evaluation schema was either flawed or unfair towards non-North Americans (as what a higher "quality" abstract means could be somewhat subjective), or the organizers did not do an adequate job to inform the evaluators of potential bias. Thus, it is possible that the style and/or structure of an abstract that is generally considered "good" by researchers in North America (in terms of the four evaluation categories used by AbGradCon 2021: merit, literacy, novelty, and soundness) is different from those considered "good" by researchers in other regions. As such, it is possible that the North American evaluators evaluated abstracts with the expectation that a "good" abstract would be written in the "North American style", which was not familiar to those in other countries.

While the organizers were explicit in informing the evaluators that English ability of the applicant based on perceptions from the abstract was not to be judged so that the conference would be accessible to a wide-range of English levels (as long as the participant could communicate effectively in English, they could attend the meeting), it is possible that English ability also affected the ability for abstract evaluators to understand scientific aspects of the abstract (or, evaluators from North America were not used to evaluating such abstracts, and gave lower scores subsequently despite such abstracts having no problems in terms of scientific content). In particular, while English education differs greatly by country, region, and even within each country, a number of countries in Asia do not have strict English language requirements or many English-based schools or educational opportunities *en masse* (as compared to those in North America, Europe, or Oceania), leading to the possibility of less opportunities and experience for applicants from Asia to practice or participate in English language-based scientific exchange, programs, or events, which may have been reflected in these abstract evaluations. However, this is simply speculation. Nevertheless, we will note again that although we believe that the evaluation process was fair, the evaluator selection process was not rigorous or systematic, which may have resulted in some of the biases observed herein.

**Table 6. Statistics of All Submitted Abstracts (mean score of all evaluated abstracts was 12.1). A heteroscedastic two-tailed t-test was performed comparing the mean abstract evaluation scores of the abstracts in the category listed in the same row, against mean abstract evaluation scores of all other abstracts. Highlighted values are of p<0.05, indicating statistical significance.**

| Region | Mean Evaluation Score | Standard Deviation of Evaluation Score | Number of Abstracts | Heteroscedastic t-Test p-value |
|---|---|---|---|---|
| Asia | 11.3 | 2.4 | 44 | 0.00812694 |
| Europe | 12.2 | 1.5 | 40 | 0.76639675 |
| Middle East | 12.8 | 2.8 | 2 | 0.8063709 |
| North America | 12.4 | 1.8 | 117 | 0.01441121 |
| Oceania | 12.7 | 2.4 | 7 | 0.51970123 |
| South America | 11.7 | 1.9 | 15 | 0.5456408 |

| Primary Field | Mean Evaluation Score | Standard Deviation of Evaluation Score | Number of Abstracts | Heteroscedastic t-Test p-value |
|---|---|---|---|---|
| Astronomy | 12.2 | 1.1 | 16 | 0.91303232 |
| Biology | 11.9 | 2.2 | 88 | 0.27455093 |
| Chemistry | 12.9 | 1.5 | 39 | 0.00143995 |
| Education, Outreach, Diversity | 12.2 | 0.8 | 3 | 0.85266504 |
| Engineering and Instrument Design | 12.0 | 1.1 | 5 | 0.86566722 |
| Geology | 12.2 | 1.7 | 20 | 0.75576246 |
| Philosophy of Science | 7.6 | 0.8 | 2 | 0.07180028 |
| Physics | 12.5 | 2.3 | 4 | 0.78175869 |
| Policy | 8.4 | 2.9 | 2 | 0.32232786 |
| Planetary Science | 12.3 | 1.6 | 32 | 0.60505461 |
| Other | 11.7 | 2.3 | 14 | 0.5052973 |

*4.4 Abstract Acceptance/Selection*

Of the applications, 146 abstracts were accepted to AbGradCon 2021 on the basis of their scores and other factors. First, as is customary for traditional in-person AbGradCon meetings, local attendees who submit a valid and relevant abstract are accepted regardless of score, and as such, all participants based in Japan were accepted to the meeting regardless of score. Additionally, all applicants who contributed to organizing the meeting (including discussion leaders and abstract evaluators) were accepted regardless of score. Finally, those who were co-winners or honorable mentions from the AbGradCon 2020 Undergraduate Flash Talk Contest were also accepted.

Next, we accepted abstracts based on evaluation scores. 115 applicants with a final abstract score of 12.4 (out of a maximum 16 points) or higher were accepted. However, this resulted in an overly skewed representation of the meeting towards participants from North America (and in particular, the USA), with 62 of the 136 accepted abstracts (46%) coming from those based in the USA. As the theme of the meeting was international collaboration, further increasing the diversity of participants (based on country of residence) was one very valid way to increase the potential for international collaboration; a meeting with almost half of the participants based in a single country would not have been conducive towards such goals. As such, all non-USA-based applicants with a final abstract score of 11.67 or above (which was 10 applicants) were also accepted to the meeting, leading to the final total number of accepted abstracts.

As such, of the 146 abstracts accepted, the mean score was 13.1 (standard deviation of 1.3), while the minimum score was 6.75; the maximum score was 15.67, as reported earlier. Of the abstracts submitted by organizers and of those based in Japan, 11 organizers and six participants based in Japan would not have been accepted had we not guaranteed attendance to these groups of people. Ultimately, 119 of these accepted participants registered for the meeting; those who withdrew cited different reasons for withdrawal, including conflicts with other events, conflicts with regular work/research/school duties, conflicts with field excursions, etc.

*4.5 Live-streamed presentation selection*

The selection of live-streamed presentation speakers was the result of sequential elimination based on criteria recorded from the participant questionnaire answers received. In the case of an in-person meeting, the oral presentations would have been chosen first by field (Biology, Geology, Planetary Science, Astronomy & Physics, Chemistry, or Engineering & Instrument Design) followed by the individual participant's ranking in the abstract evaluation. Some correction was planned to more equally divide speakers by country of residence to increase the international representation of the meeting. In the case of a digital meeting (as in AbGradCon2021), these criteria were still the general basis for selection but additional parameters needed to be included. Questions in the participant questionnaire included information about whether or not a participant would prefer to prepare a full oral presentation, a flash talk, or a combination of the two. Those who selected a flash talk alone were removed from the pool of potential live speakers. Second, we asked participants if they wished to be removed from consideration for a live presentation, yielding the opportunity to a different participant. Anyone who yielded their spot was removed from the pool of potential options.

The next selection criteria was specific to an online meeting across different time zones: availability. In a traditional AbGradCon event, all of the participants would in theory be present at all times during the meeting. With the online format, we had both a morning and an afternoon session (in JST) to account for differences in time zones. Live-streamed speakers were expected to answer questions live during their session and would be required to attend the session for this reason. As such, we included one question in the participant questionnaire which asked about the participant's availability for each session on each day of the conference classified as: in full, part of the time, or none of the time. This required eight individual questions (eight sessions over four days) within the questionnaire to guarantee that the information was sorted in a useful way within the master spreadsheet. Initially, those participants who could not attend on the days selected for talks within their scientific field were removed, but exceptions were later made to include certain participants who replied "part of the time". This selection required the most amount of careful consideration and time to make certain that enough speakers were available for the various times by scientific category, and the program was changed to account for these differences in availability for fields which had far fewer participants available. An email was sent to all the selected live-streamed presenter candidates to confirm that they were still willing to provide a live-streamed presentation before the final programme was confirmed. Those who decided to withdraw at this stage were replaced by alternate candidates following the above mentioned process and speakers were notified of their assigned presentation times. In the end, the morning sessions were populated primarily by participants in the North and South Americas with inclusion of a few participants in Asia. Collectively, many Asia-based participants had declined to be considered for the live-stream session. The afternoon session was primarily speakers from Europe and Asia, with occasional North American speakers who had selected to remain online well-outside of the "normal" working-times. The selection process itself relied heavily on carefully crafted questions in the participant questionnaire which allowed for the determination of viable live-streaming candidates for the required times of the meeting. A total of 36 different speakers were selected by this method to present in 12 different live sections divided into six different scientific fields of concentration. Each of the six scientific fields was represented twice; once in the morning session and again in the afternoon, allowing for different speakers in both session time zones.

## 5. Conclusions and Recommendations for Future Meetings

Throughout the AbGradCon 2021 organization and execution process, a number of lessons were learned by the organizers with respect to conference organization and execution, while a post-conference survey (39 responses) also highlighted a few key areas of need for future conferences. For example, even though most of the respondents were satisfied with the execution of the meeting, only about half of survey respondents believed that a virtual AbGradCon was a sufficient substitute for an in-person conference, as interacting and networking in-person at conferences simply cannot be replaced by 100% virtual interactions. Similarly, while the career building events/activities that were organized for 2021 (proposal review panel, science communication workshops, etc.) were generally highly rated and most participants felt that they were useful and worth attending, respondents requested not only more career building activities, but also events/activities focused on outreach, diversity and inclusivity, as well as personal/mental health in future meetings. Finally, the post-conference survey response rate was around 33% and is quite low compared to previous in-person AbGradCon meetings, some of which had response rates approaching or exceeding 90% of participants. This could be attributed to participants generally feeling less “involved” in the actual conference or that a smaller number of attendees participated in the conference “fully” and felt like they had something to contribute to such a survey. Combining the lessons learned from organizing and executing this conference with the survey results leads us to issue a number of recommendations for future meetings regarding virtual meeting organization, internationality of meetings, and interdisciplinarity in meetings in hopes of supporting and being useful to future virtual meetings in astrobiology and other fields.

### *5.1 Virtual Meeting Organization*

- For virtual meetings where not all attendees can join all sessions due to time zone reasons, asynchronous communication is necessary to build connections and networks amongst the entire attendee population.

We observed that the platforms that we used, Gather.town and Discord, were able to provide space for asynchronous meetings and discussions amongst the AbGradCon 2021 participants, most of which were self-organized. In particular, some of these asynchronous discussions may have led to formation of new collaborations, while one outcome of these discussions that emerged naturally was the formation of a group of attendees interested in outreach and education, which resulted in a session proposal application for a future conference (AbSciCon 2022). However, in order to accomplish this, the contact information of participants must be listed and categorized clearly for ease of searching and identification by other attendees (to initiate discussions), something that was accomplished through our conference booklet.

- On-demand pre-recorded presentation videos can allow participants to view videos at their leisure, while streaming pre-recorded videos during plenary sessions can lead to a smoother conference with less technical issues.

We believe that providing the presentation videos to all participants to view at any time allowed many of the participants to participate at their leisure (in case they could not participate live during any conference sessions). This is important as the differences in time zones as well as in personal education/research schedules may not allow full attendance to all sessions. Additionally, this may spur further discussion between attendees. Finally, publishing the presentation videos of participants who agree to public release (the collaborators and supervisor(s) of these participants must also agree) could lead to more feedback to the participants from the astrobiology community at-large, while also helping to increase the exposure of early career researchers to the public (for example, a sharable video link) and to other researchers, an important aspect to further one's scientific career. Finally, while streaming pre-recorded videos puts much more strain on the organizing team to collect, organize, upload, and stream the videos, it also leads to a much more streamlined conference with less chance of derailment or delayed sessions due to technical issues of speakers, some of whom may not have stable internet connections.

- Conference sessions should not be more than two or three hours long at one time.

In order for the participants to maintain focus on the conference, it appears that virtual conference sessions work most effectively when such sessions are less than three hours long. Due to the amount of content, if necessary, a planned long conference session (six to eight hours) could be broken up into two or three shorter sessions and spread out throughout the day with ample breaks; for example, a break of at least two hours between sessions.

*5.2 Internationality of Meetings*

- Sessions of international conferences which are held at times which allow attendance of people from a variety of locations help to improve diversity with respect to internationality and may also encourage international collaborations.

Digital meetings offer an opportunity for a number of early career researchers and students who cannot easily attend in-person meetings, due to travel demands or lack of funds, to participate in scientific activities. This is especially true for participants from many countries in the Asia region which may not support scientific activities in astrobiology (especially for early career researchers) to the same level as other countries. As such, in order to increase the diversity of participants in an international virtual conference, the time zones of sessions should be designed to maximize participation from a variety of locations. For example, AbGradCon organized sessions during times which allowed most participants from East Asia to join all sessions, while also meeting their peers from North America and Europe. We notice that many virtual conferences, especially in astrobiology, are held at times which do not allow easy attendance by those in Asia and Oceania, and thus more conferences which are held at times which allow such attendance are absolutely necessary in order to increase participation of researchers in these regions, which are attempting to grow their astrobiology research footprint.

- The abstract submission and evaluation process should be designed so that submissions from all regions are treated equally and fairly.

We noticed in our analysis of the AbGradCon 2021 abstract evaluations that there may have been some regional bias in terms of evaluation scores. Whether this bias was due to the actual higher "quality" of North American abstracts or due to a somewhat biased evaluation process towards non-North Americans must be considered. If the former, then more information about what a "good" abstract looks like and how the abstract will be evaluated should be provided to all potential applicants and evaluators to increase fairness across applications from all regions, while if the latter, clearly defined expectations and definitions must be provided to the evaluators to prevent such regional bias. We implemented completely blind evaluation (none of the authors or co-authors of any abstracts were known to any evaluators), which we recommend to future conferences. We also recommend that more evaluators from around the world, including more from regions outside of North America, be recruited for any abstract evaluation processes.

*5.3 Interdisciplinarity in Meetings*

- Implement abstract evaluations across fields, rather than limited just to the field of the respective evaluators.

We found that there were generally (with some minor exceptions) no significant differences in abstract evaluation scores across different disciplines, suggesting that evaluation of abstracts of any field by evaluators of many fields in our case likely resulted in a fair evaluation process with respect to scientific discipline. Thus, for conferences with broad scientific topics included (as was the case for our conference), we recommend that abstract evaluations are similarly implemented to allow evaluations of abstracts not only within one's own field of expertise, but also of abstracts in other fields' relevant to the conference topic. This may help evaluators gain a more holistic understanding of all topics covered in the conference, while also selecting for abstracts that can explain the content and important concepts to those from a variety of fields (something crucial to researchers in an interdisciplinary field such as astrobiology).

- Breakout sessions combining participants from a variety of disciplines, along with discussion leaders from a variety of disciplines that can facilitate participation by all participants, can help increase interdisciplinarity of a meeting.

In the AbGradCon 2021 breakout sessions, all of the participants were distributed randomly irrespective of field. This provided a forum for participants to interact directly in smaller groups with those from other fields, leading to more interdisciplinary discussions. Additionally, the discussion leaders' were tasked with encouraging participation from all in their discussion session;

as direct participation may increase understanding or at least interest in a topic, this may have led to some participants exploring ideas or topics in a field different from their own. Finally, we had the fortune of having discussion leaders from a variety of fields. If participants participated in all breakout sessions throughout the entire meeting, then they were likely exposed to viewpoints and ideas from not only a variety of participants, but also from discussion leaders from different fields. As many conferences are focused on a very specific topic, it may be difficult for participants to join interdisciplinary discussions in such fora, and thus directed and deliberate organization of discussions (likely by the organizers or by discussion leaders) are necessary to increase interdisciplinarity in discussions. In particular, some of the breakout sessions resulted in proposals for other astrobiology-related discussion topics and questions that should be discussed and answered by the community as a whole.

- Speakers and sessions from a variety of disciplines should be featured during a plenary session, and such presentations should not be heavily skewed towards one particular discipline or another.

The AbGradCon 2021 plenary sessions (live-streamed pre-recorded presentations) covered a variety of topics in roughly equal proportion. This allowed participants to be exposed to topics other than their own; such exposure to different topics is very unique to the astrobiology field (and other fields which are inherently interdisciplinary) and is essential for the career development of early career astrobiology students and researchers. AbGradCon 2021 was a single-track conference with no overlapping plenary sessions, which meant that participants who decided to attend all (or most) plenary sessions heard presentations from much more than their own field. For larger conferences with more than one "track" of plenary sessions, it is possible for one to attend sessions at all times offered and still only be exposed to presentations of one's own discipline. As such, we highly recommend smaller conferences which can accommodate a single track (for larger conferences, this is not possible, and we acknowledge the difference in goals between such conferences) to contain presentations from a variety of fields.

*5.4 Prospective*

In the case of AbGradCon 2021, we hoped to highlight and encourage interdisciplinarity and international collaboration in astrobiology research, and we hope that we accomplished our goals; we can neither measure the success objectively nor over short timescales, and thus we must wait more years for the attendees to mature as researchers and start their independent research careers. While other conferences may have different goals, we hope that some of the lessons and information learned through organizing and executing virtual AbGradCon 2021, and the subsequent recommendations, will be helpful for future conference organizers within and outside of the astrobiology field.

**Author Contributions**

All authors wrote and edited the manuscript and approved the final version.

**Competing Financial Interests**

No competing financial interest declared.

**Acknowledgements**

We would like to thank all of the sponsors and partners of AbGradCon 2021 for their continued support: Earth-Life Science Institute (ELSI) of the Tokyo Institute of Technology, the NASA Astrobiology Program, Blue Marble Space Institute of Science (BMSIS), Japan Society for the Promotion of Science (JSPS) Grant-in-Aid for Scientific Research on Innovative Areas Aquaplanetology (17H06454), the Astrobiology Center (ABC) Program of Japan National Institutes of Natural Sciences (NINS), the ELSI-FirstLogic Astrobiology Donation Program (EFADP), Astrobiology Australasian Meeting 2020, AbGradE, Explainables, Astrobiology Club, Astrobiology Society of Asia-Pacific (ASAP), *Life*, Breakthrough Listen, Greenspace, and SSOEL: Astrobiology-Japan. We would further like to thank the keynote speakers H. Yabuta (Hiroshima University) and A. Antunes (State Key Laboratory of Lunar and Planetary Sciences at Macau University of Science and Technology (MUST)), the Explainables Team (D. Angerhausen, A. Salleh, J. John), M. Neveu (NASA Goddard Space Flight Center), M. Kirven-Brooks (NASA), A. Gronstal (NASA), M. Toillion (NASA), A. Sato (ELSI), H. Ricciardi (ELSI), H. Tanaka (ELSI), K. Akiyama (ELSI), R. Hattori (ELSI), L. Kwok (ELSI), S. Som and J. Haqq-Misra of BMSIS, D. Klock (Greenspace), J. DeMarines (Breakthrough Listen), K. Aguilar (Breakthrough Listen), A. Siemion (Breakthrough Listen), and the various discussion leaders of the conference (A. Mojarro, K. Matange, M. Naseem, D. Phillips, K. Miller, R. Mizuuchi, M. Macey, S. Sarkar, S. Biswas, M. Schaible, R. Archer, S. Borges, G. Schaible, A. Deal, L. Schwander, Y.A. Takagi, C. Walton, A. Kahana, S. Choi, D. Buckner, B. Kang, Y. He, S. Nishikawa, S. Sharma, and A. Verma). Finally, a special thanks to the other organizers who could not participate in this article, including K.W. Lim (ELSI), W. Takahagi (University of Tokyo), L. Vincent (University of Wisconsin-Madison), T. Roche (Georgia Tech), S. Vissapragada (Caltech), etc., as well as the organizers of AbGradCon 2022.

T.Z.J., K.N.J.-F., K.F., T.H., Y.L., N.N., P.P., and H.B.S. are members (and I.B. is a former member) of ELSI, which is sponsored by a grant from the Japan Ministry of Education, Culture, Sports, Science and Technology (MEXT) as part of the World Premier International Research Center Initiative (WPI). During the period of organization of AbGradCon: T.Z.J. was further supported by JSPS Grants-in-aid 18K14354 and 21K14746, the Tokyo Institute of Technology Yoshinori Ohsumi Fund for Fundamental Research, project grants from the Astrobiology Center Program of Japan National Institutes of Natural Sciences (NINS) (Grant Numbers AB311021, AB021008), and Tokyo Institute of Technology Seed Grant "Tane" 1798; I.B. was further supported by a Ministry of Education, Culture, Sports, Science and Technology (MEXT) Scholarship and a JSPS DC2 Fellowship for Young Scientists; N.G. was further supported by the NASA-funded Laboratory for Agnostic Biosignatures Project and the Santa Fe Institute; K.F. was further supported by the EFADP; Y.L. was further supported by JSPS Grants-in-aid 20H04608 and 19K15671; and T.Z.J. and Y.L. were both further supported by the Assistant Staffing Program by the Gender Equality Section, Diversity Promotion Office, Tokyo Institute of Technology.